\documentclass{aa}
\usepackage{graphicx}
\usepackage{txfonts}

\begin{document}
\title{On highly eccentric stellar trajectories interacting\\
with a self-gravitating disc in Sgr~A$^{\star}$}
\author{L.~\v{S}ubr\inst{\!1} \and V.~Karas\inst{2,1}}
\institute{$^1$~Charles University, Faculty of Mathematics and Physics,
 V~Hole\v{s}ovi\v{c}k\'ach~2, CZ-180\,00~Prague, Czech Republic\\
$^2$~Astronomical Institute, Academy of Sciences, Bo\v{c}n\'{\i}~II, 
 CZ-141\,31~Prague, Czech Republic}

\authorrunning{L.~\v{S}ubr \& V.~Karas}
\titlerunning{On highly eccentric stellar trajectories in Sgr~A$^\star$}
\date{Received 30 September 2004; accepted 30 December 2004}
\abstract{We propose that Kozai's phenomenon is responsible for the long-term 
evolution of stellar orbits near a  supermassive black hole. We pursue
the idea that this process may be driven by a fossil accretion disc in
the centre of our Galaxy, causing the gradual orbital decay of stellar
trajectories, while setting some stars on highly elliptic orbits. 
We evolve model orbits that undergo repetitive transitions
across the disc over the period of $\approx10^7$ years. We assume that 
the disc mass is small compared to the central black hole, and its
gravitational field comparatively weak, yet non-zero, and we set the
present values of orbital parameters of the model star consistent with
those reported for the S2 star in Sagittarius~A$^{\ast}$. We show how 
a model trajectory decays and circularizes, but at some point the mean
eccentricity is substantially increased by Kozai's resonance. In
consequence the orbital decay of highly eccentric orbits is
accelerated. A combination of an axially symmetric gravitational field
and dissipative environment can provide a mechanism explaining the
origin of stars on highly eccentric orbits tightly bound to the central
black hole. In the context of other S-stars, we can conclude
that an acceptable mass of the disc (i.e., $M_\mathrm{d}\lesssim1$~per
cent of the black hole mass) is compatible with their surprisingly young
age and small pericentre distances, provided these
stars were formed at $r\lesssim10^5$ gravitational radii.
}
\maketitle

\section{Introduction}  
A dense nuclear cluster surrounds the centre of our Galaxy.
Tracking the rapid proper motion of individual stars in the central
arcsecond (i.e., about $0.04$~pc off Sgr~A$^{\ast}$ object) provides 
an essential tool to determine the central mass
$M_{\bullet}\approx3.5\times10^6M_{\sun}$ (Genzel et al.\ 
\cite{genzel03}; Ghez et al.\ \cite{ghez03a}). So far, the motion of S2 
(S0-2) has been measured with the best accuracy over a sufficient
duration of time (Sch\"odel et al.\ \cite{schodel02}, \cite{schodel03}).
It is a highly eccentric, $e=0.87$ Keplerian orbit with the orbital
period of $15.2$ years and an apocentre distance $0.01$~pc. Other newly
discovered stars strengthen the case for a supermassive black hole in
Sgr~A$^{\ast}$, and for the surprisingly young stellar population around
it. The rather high mass and low age present a challenge to star-formation
theories (Milosavljevi\'c \& Loeb
\cite{mil04}; Eckart et al.\ \cite{eckart04}). In this paper 
we discuss one of the possible mechanisms that could help to resolve
the issue.

S2 is a member of the central Galactic cluster for which radial velocity
as well as proper motion have been reported by Ghez et al.\
(\cite{ghez03b}). Combination of the two pieces of information allowed
these authors to determine the orbital parameters from the full
three-dimensional motion and to break the ambiguity in  the inclination
angle. Trajectories and spectroscopic measurements  of other stars in
this region have not yet reached comparable accuracy, however, there are
good prospects for future progress and strong indications of the puzzling
situation: a number of S-stars are on close orbits near the black hole
but they seem to be of young age, contrary to expectations. Several
mechanisms have been proposed that could bring stars to the
neighbourhood of the central black hole (e.g., Hansen \& Milosavljevi\'c
\cite{han03}; McMillan \& Portegies Zwart \cite{mcm03}; Alexander \&
Livio \cite{alex04}). In spite of this effort the origin of close stars
is not well understood. Models have difficulties in reconciling different
aspects of the Galaxy centre -- on the one hand it is a low level of
present activity, indicating a very small accretion rate, and on the
other hand the spectral classification suggests that these stars have
been formed relatively recently. While in-situ star formation is
problematic because of strong tidal forces operating near the central
black hole, two-body gravitational relaxation could in principle set
stars on plunging orbits. Nonetheless, the latter mechanism does not
seem to be efficient enough because a rough estimate of the relaxation
time-scale, $t_\mathrm{r}\gtrsim1$~Gyr, is orders of magnitude longer
than the estimated age of the S-stars ($\approx 10$~Myr). We
neglect the two-body gravitational scattering and the effect of
gravitational relaxation in the present analysis (see Alexander \&
Hopman \cite{alex03} for a more elaborate discussion of gravitational
relaxation in the Galactic Centre).

The interaction between stars and a fossil accretion disc is a
cumulative effect which provides a viable mechanism of transporting
stars towards the black hole, while stripping their outer atmosphere
away. These two effects are mutually linked and they
complement one another: whereas the stellar orbits are evolved by the
interaction with the disc, both gravitational and hydrodynamical
as discussed in this paper, the stars
themselves as well as the composition of the gaseous environment are
also affected (Artymowicz et al.\ \cite{art93b}; Armitage et al.\
\cite{arm96}). The main appeal of this scheme lies in the fact that it
naturally combines several influences which are needed in order to
explain the efficient migration of stars to the centre: the proposed
scenario sets some stars from the sample on elliptic orbits for a fraction
of their lifetime (the gravitational effect of the disc), it also causes
continuous migration (hydrodynamical effect of the orbital dissipation),
and finally it affects stars by removing their atmospheres. All these
effects are determined by the assumed density profile of the disc and
they are thus interconnected. Furthermore, it
was proposed that S-stars have undergone rejuvenation by the interaction
with the environment or by tidal squeezing during close encounters with the
black hole, thereby hampering the spectral classification and 
mimicking young age (e.g.\ Morris \cite{morris93}; Alexander \& Livio
\cite{alex01}). Notice, however, that we neglect these peculiarities,
nor do we take into account the likely effect of the disc
itself evolving.

Various authors have investigated the long-term orbital evolution of a
stellar satellite that is either periodically crashing on the black-hole
accretion disc (e.g., Syer et al.\ \cite{syer91}; Vokrouhlick\'y \& Karas
\cite{vok93,vok98}; Rauch \cite{rauch95};
\v{S}ubr \& Karas \cite{subr99}) or stays embedded in the disc plane
(Artymowicz \cite{art93}; Hameury et al.\ \cite{ham94}; Karas \&
\v{S}ubr \cite{karas01}). The above-mentioned papers address the
star--disc interaction in active galactic nuclei, whose accretion discs
are supposed to be rather dense, however, it has been suggested (Levin
\& Beloborodov \cite{levin03}; Nayakshin \& Sunyaev \cite{nay03}) that
also our Galaxy centre contained a similar gaseous disc at some stage of
its history. Some form of an inactive disc or a dusty torus may still
exist as a remnant of past activity. If such a disc existed in the
Galaxy centre, it could capture passing stars from unbound orbits, or it
could directly supply the medium where new stars form and evolve while
sinking to the black hole. This mechanism has been explored under
conditions relevant for active galactic nuclei and quasars (Collin \&
Zahn \cite{collin99}; Goodman \& Tan \cite{goodman04}) and for
Sgr~A$^\star$ (Genzel et al.\ \cite{genzel03}). Likewise it has been
speculated that star-disc collisions may contribute to the featureless
X-ray variability by creating bright spots on the disc surface and
fountains of ejected material above it (Zentsova \cite{zen83}; Karas \&
Vokrouhlick\'y \cite{karas94}; Vilkoviskij \& Czerny
\cite{vilk02}). Cuadra et al.\ (\cite{cuadra03}) and Nayakshin et al.\
(\cite{nay03}, \cite{nay04b}) have pointed out that a similar mechanism
may be responsible for X-ray flares, the occurrence of which was first
reported  by Baganoff et al.\ (\cite{bag01}). A sufficiently long
observation could reveal periodicities of the stellar motion connected
to precession of the orbit, and from them one would be able to infer
the angular momentum of the central black hole (Semer\'ak et al.\
\cite{sem99}; Aschenbach et al.\ \cite{asch04}).

We apply the model of star--disc collisions to the orbital
evolution of the star S2, assuming a dissipative perturbation of its motion
by crossing an inactive gaseous slab. We employ a similar formulation of
the model as described by \v{S}ubr \& Karas (\cite{subr99}) and Karas
\& \v{S}ubr (\cite{karas01}). The relative mass of the disc, 
$\epsilon\,\equiv\,M_\mathrm{d}/M_{\bullet}$, is a free parameter for us. We
adopt a plausible and quite conservative value of $\epsilon$ in the
range $10^{-3}\la\epsilon\la10^{-2}$ (see also Nayakshin \& Cuadra
\cite{nay04a}; \v{S}ubr et al.\ \cite{subr04}). 
Simulations are initialized at $r_\mathrm{i}\gtrsim3\times 10^4R_\mathrm{g}$ 
(i.e.\ above the present S2 distance; 
$R_\mathrm{g}\,\equiv\,GM_{\bullet}/c^2\approx1.9\times 10^{-7}
\mathrm{pc}$) and we follow the sinking satellite as it proceeds to
the centre. The gas-assisted drag tends to circularize the orbit and
align it with the disc plane, while the Kozai (\cite{koi62}) mechanism
drives eccentricity to a large value, characteristic of the present
stage of S2. We employ this mechanism to explain the origin of the large
orbital eccentricity and small pericentre. The assumed mass of gas
distributed in the disc and the resulting perturbation of the predicted
orbits are small enough to be consistent with Keplerian motion over one
(or several) revolutions, which is the time span accessible to direct
observational checks (see Mouawad et al.\ \cite{mou04} for an upper
estimate of the mass outside the black hole). Clearly visible departures
from perfectly elliptical orbits are expected on a substantially longer
time-scale, $\approx10^3$ orbits. 

\section{The orbit evolution}
Let the gravitational field of the central black hole be described in
the Newtonian approximation and the disc be planar and geometrically
thin. Passages of a star through the disc last only a small fraction of
the orbital period at the corresponding radius, and so they can be
considered as instantaneous events at which the passing body exchanges
momentum with the material lying along its trajectory. Repetitive events
cause the orbit to decay. Outside the disc plane the star moves freely
and its orbit is found numerically by integration in the combined
gravitational field of the central mass $M_{\bullet}$ and the disc mass
$M_\mathrm{d}$. This requires one to specify the surface density
distribution as a function of radius in the disc and we  assumed two
cases for simplicity. First we employ a homogeneous slab
$\Sigma_\mathrm{d}(R)=\mathrm{const}$. Its gravitational field can be
then expressed analytically (e.g., Lass \& Blitzer \cite{lb83}; Pierens
\& Hur\'e \cite{pierens04}), which helps to speed up computations.
Another variant,  $\Sigma_\mathrm{d}(R)\,\propto\,\sqrt{R(R_\mathrm{out}
- R)}$,  was used to mimic the surface density and the gravitational
field of an  annulus ($R_\mathrm{in}{\leq}R{\leq}R_\mathrm{out}$). Given
the value of $M_\mathrm{d}$, the non-sphericity of the gravitational
potential will be smaller and its impact on stellar orbits weaker if the
disc matter is concentrated more towards the centre. However, we checked
that a very similar effect to that  described here is obtained even with
the standard Shakura--Sunyaev  density profile, i.e.\ decreasing with
radius as $\Sigma_\mathrm{d}(R)\,\propto\,R^{-3/5}$. Most likely the
actual density distribution will be somewhat different because the disc
will be modified by self-gravity, stellar transitions and other 
effects.\footnote{By the effects of {\em self-gravity} we understand the
gravitational influence of the disc on the stars. As mentioned above, 
we do not modify the internal structure of the disc itself although this
may also be interesting to examine because self-gravity is expected to
cause a clumpy structure of the disc, thereby reducing the efficiency of
star--disc collisions.}

The assumptions about the form of the disc are necessary for
definiteness of examples, but they do not affect the nature of our
results. If the trajectory is oriented in such a way that it crosses the
disc repetitively, we observe the gradual dissipation of orbital energy
and of angular momentum. As time proceeds, the orbit becomes circular
and declined in the disc plane, but this overall trend is overlayed with
oscillations of eccentricity and inclination. Occasional jumps toward
high eccentricity occur due to self-gravity of the disc, acting as a
perturbation in our model.

\begin{figure*}
\includegraphics[width=\textwidth]{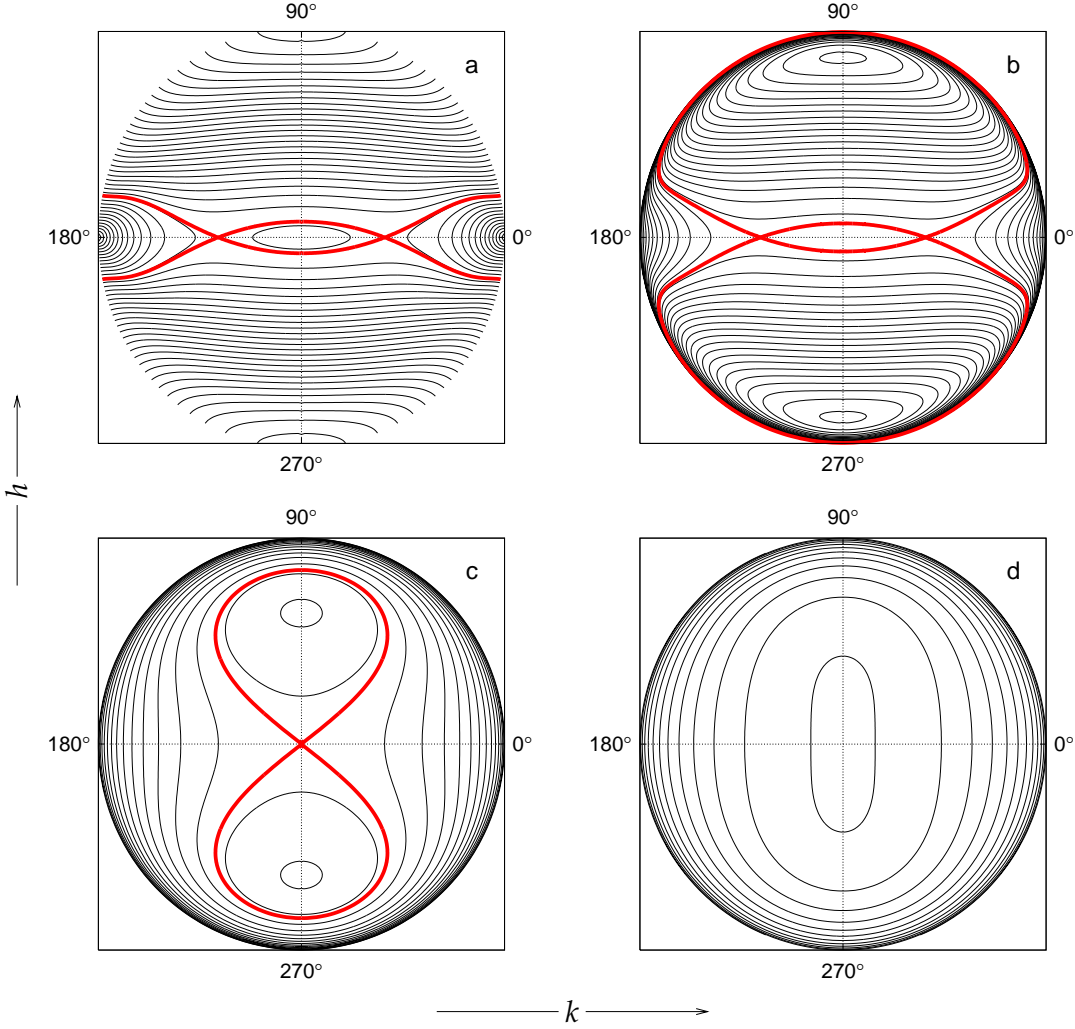}
\caption{Contour lines of the disturbing function
$\bar{V}_1(\rho,\omega)$ (shown with equidistant steps between contour 
levels). The polar plot is shown with zero eccentricity at the origin
$\rho=0$ ($e=0$) and polar angle $\omega$ indicated on margins. In the
case of negligible dissipation the orbital evolution proceeds along
contours. In different regions of the graph $\omega$ either circulates
over the full circle or it librates within a limited interval. A
separatrix (thick line) forms a border between these regions. Four
topologically different cases can be distinguished by the structure of
contours, shape of the separatrix and the corresponding value of Kozai
constant $c$: (a)~$c=0$ ($e<1$); (b)~$c=0.2$ ($e<0.98$); (c)~$c=0.8$
($e<0.60$); (d)~$c=0.9$ ($e<0.44$). These define the range of orbital
oscillations of eccentricity and inclination according to
eq.~(\ref{kozai}). Orbital energy has been set identical and equal to
$-GM_{\bullet}/(2a)$ with semi-axis $a=2.4\times10^4R_\mathrm{g}$ in all
panels. While on the medium time-scale the orbits remain attached to
underlying contours, allowing us to estimate the range of variations of
eccentricity, on the long time-scale  the orbits are driven across
contours. This gradual diffusion leads to topology changes at some
moments.}
\label{fig0}
\end{figure*}

Kozai's method was developed to describe the evolution of mean
orbital elements of a small body disturbed by the presence of another
mass in the system. Our aim here is to consider a restricted problem in
which the origin of perturbation is an accretion disc and the star
represents a test body (a hierarchical triple system; see Lidov \&
Ziglin \cite{lidov76}; Thomas \& Morbidelli \cite{thomas96}; Kinoshita
\& Nagai \cite{kino99}). Kozai's resonance is a particular type of
secular resonance occurring when the pericentre argument precession
vanishes, $\dot{\omega}=0$ for the test body. In order to disregard
various short-time effects acting on dynamical time-scale, the method
invokes averaging technique which is based on canonical
transformation (Arnold \cite{arnold89}; Brower \& Clemence
\cite{brower61}; Morbidelli \cite{morbi02}). Although this formalism was
suited and intensely applied to the problem of motion in the Solar
system, it has met numerous applications elsewhere. For example, Holman
et al.\ (\cite{holman97}) proposed that large eccentricities of the
orbits of some extrasolar planets could be understood in terms of Kozai
oscillations, which are caused by the disturbing influence of the secondary
star. A similar approach was employed in the context of stellar dynamics
around massive black holes in galactic nuclei (Sridhar
\& Touma \cite{sridhar99}) and in globular clusters (Wen \cite{wen03}).
Blaes et al.\ (\cite{blaes02}) proposed that oscillations of
eccentricity could be important for determining the rate of binary black
hole merging and gravitational wave emission expected from galactic
centres.

An important point to notice is the presence of dissipation due to stars
crossing the disc. The star experiences orbital decay due to repetitive
transitions changing the trajectory and pushing it across the
separatrix. We consider two effects which both depend on the density of
the gaseous environment: the energy dissipation is roughly proportional
to the density of the medium and likewise the  gravitational
perturbation increases its impact proportionally to the mass contained
in the disc. These two processes are therefore mutually connected. As
far as star--disc collisions  are concerned, relevance of the
Kozai--type mechanism coupled with dissipation was originally discussed
by Vokrouhlick\'y \& Karas (\cite{vok98b}) and we refer the reader to
their paper for further details. Here we briefly summarize the adopted
approach.

The origin of the disturbing gravitational influence is self-gravity of
the accretion disc, hence the perturbation potential  is already assumed
axially symmetric. Averaging is performed over the star ellipse. We
denote the dominant potential of the central source as 
$V_\mathrm{c}(r)=-GM_\bullet/r$ (the central black hole) and the
disturbing component $V_\mathrm{d}(R,z)$ (the disc). Test-particle
motion in the referential background field is described by an
unperturbed Hamiltonian, which is simply
$\mathcal{H}_\mathrm{c}=-\frac{1}{2}GM_{\bullet}a^{-1}$ in terms of
semi-axis $a$. Eccentricity and inclination are related by\footnote{In
our notation, $i=0\degr$ corresponds to an orbit co-rotating with the
disc, while $i=180\degr$ is for counter-rotating one.}
\begin{equation}
\eta\,\cos{i}=c,\quad\mathrm{with}\quad\eta\,\equiv\,\sqrt{1-e^2}
\label{kozai}
\end{equation}
and $c$ secularly constant (Kozai's integral).
Along $c=\mathrm{const}$ orbital oscillations take place on
substantially longer intervals than the dynamical
period at corresponding distances, defining the medium time-scale
$t_\mathrm{s}$ of the problem. A fairly accurate estimation of 
$t_\mathrm{s}$ can be obtained by replacing our disc with a narrow 
ring of radius $R_\mathrm{d}$ and the same mass $M_\mathrm{d}$ 
(e.g.\ Kiseleva et al.\ \cite{kis98}),
\begin{eqnarray}
t_\mathrm{s} & = &
  \frac{2}{3\pi}\epsilon^{-1}
  \left(\frac{R_\mathrm{d}}{a}\right)^3 t_\mathrm{p} \nonumber \\
& = & 556\, \left( \frac{\epsilon}{10^{-3}} \right)^{-1}
 \left( \frac{M_{\bullet}}{3.5\times10^6 M_{\sun}} \right)\,
  \left( \frac{R_\mathrm{d}}{a} \right)^3
  \left( \frac{a}{10^4 R_\mathrm{g}} \right)^{3/2}
  \mathrm{yr},
\label{ts}
\end{eqnarray}
where $t_\mathrm{p}(a)$ is the Keplerian orbital period and 
$R_\mathrm{d}{\rightarrow}R_\mathrm{out}$ characterizes the size of 
the disc. More precisely,
$c$ as well as $a$ would have been conserved if the energy dissipation
had been negligible. However, the orbital energy {\em is} gradually
dissipated in our system and this ensures that the star reaches the
separatrix and passes through it, i.e., the orbital elements change
their values and they eventually acquire a suitable combination. The
overall orbital decay introduces the third time-scale, $t_\mathrm{c}$,
and we assume that it is the longest time-scale relevant for our problem
(this assumption imposes an upper limit on density of the disc but one
can check that the constraint is safely met under astrophysically
realistic conditions). The total duration is obtained by integrating the
inverse rate of orbital decay from the outer edge of the disc down to
(say) tidal radius,
\begin{equation}
t_\mathrm{c}\propto\int_{R_\mathrm{out}}^{R_\mathrm{T}}
\dot{a}^{-1}\,\mathrm{d}a,
\end{equation}
where $\dot{a}<0$ depends on the surface density profile. Its magnitude
is roughly $\propto\,\Sigma(R)$,  but details are model  dependent (Syer
et al.\ \cite{syer91}; Zurek et al.\ \cite{zurek94}). Following the
arguments in Karas \& \v{S}ubr (\cite{karas01}) we estimate
\begin{eqnarray}
t_\mathrm{c} & \approx & 3\;10^7\,\Sigma_3\,\left(\frac{\epsilon}{10^{-3}}\right)^{-1}
\left( \frac{M_\bullet}{3.5\!\times\! 10^6M_{\sun}} \right)^2
\nonumber \\
 & & \times\,\left( \frac{R_\mathrm{out}}{10^4R_\mathrm{g}} \right)^2 
\left( \frac{a}{10^4R_\mathrm{g}} \right)^{3/2}\mathrm{yr}.
\label{tc}
\end{eqnarray}
Here we introduced $\Sigma_3\equiv\Sigma_{\ast}/(0.3\Sigma_{\sun})$ with
$\Sigma_{\ast}=M_\ast/S_{\ast}$ and $S_{\ast}={\pi}R_\ast^2$ (column
density, mass, cross-sectional area and radius of the star, and
analogically for the sun). $\Sigma_3\approx1$ for S2, i.e.\ a fiducial
value for young, massive stars. Clearly the period $t_\mathrm{c}$ 
is inversely proportional to $R_\ast^2$ and found
of the order of $t_\mathrm{c}\approx10^2$~Myr for typical
values of the model parameters.

\begin{figure*}[tbh]
\centering
\includegraphics[width=\textwidth]{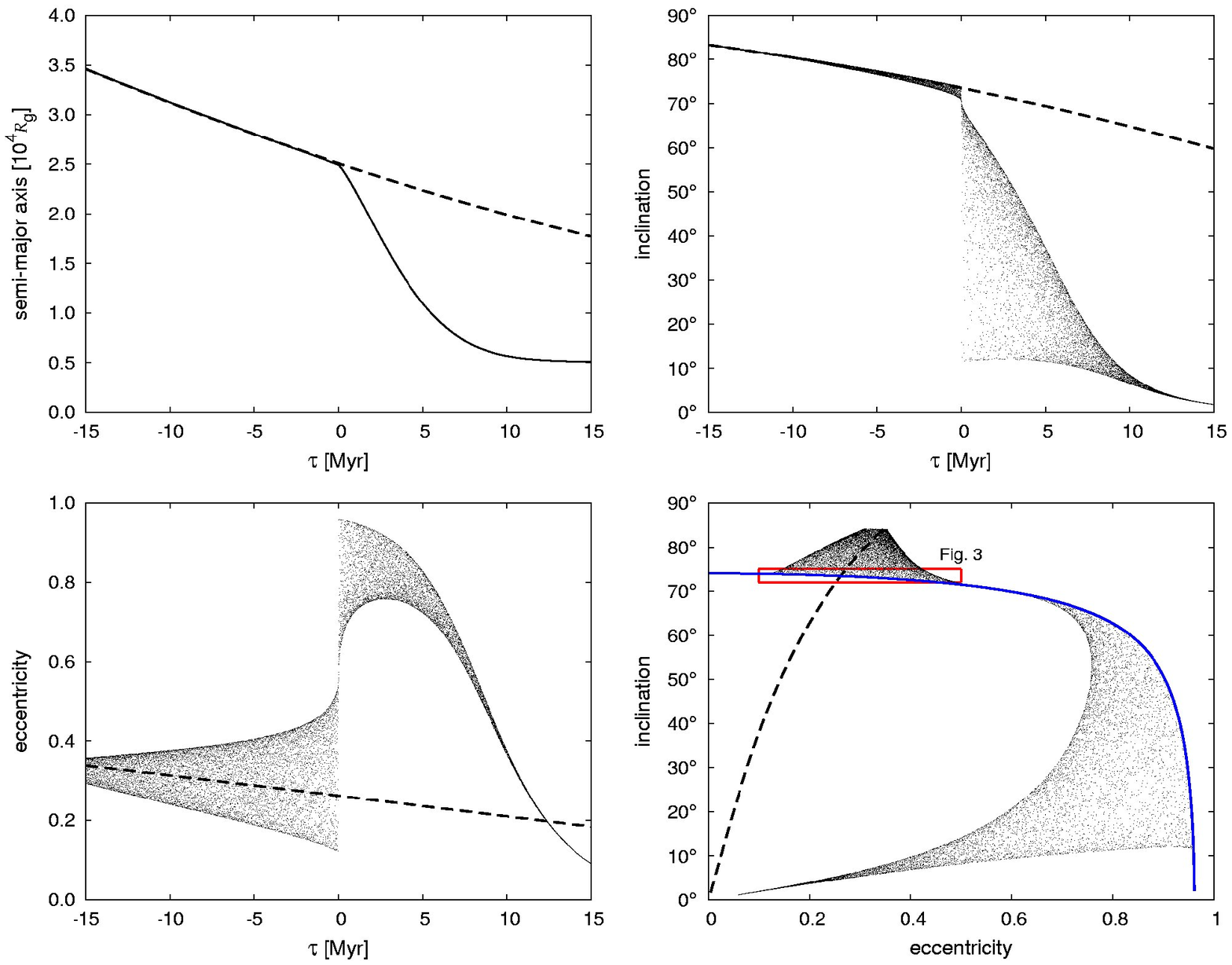}
\caption{Time evolution of a representative trajectory, 
which evolves according to the mutual interplay of gravitational and
dissipative interaction with an accretion disc. Mean orbital elements
are shown in different panels. The semi-axis $a(\tau)$ (solid line)
monotonously decays as the star sinks to the centre. Inclination
$i(\tau)$ and eccentricity $e(\tau)$ (plotted with points) oscillate with
rather large amplitudes. We also show a reference orbit (dashed line),
which was computed by neglecting the gravitational field of the disc. The
oscillations and the Kozai resonance are due to the disc gravity; hence
they affect only the exact trajectory and not the referential one. The
orbital decay is due to dissipation, hence it affects both solutions.
Both cases start with the same initial values:
$a_\mathrm{i}=3.5\times10^4R_\mathrm{g}$, $e_\mathrm{i}=0.35$ and
$i_\mathrm{i}=84\degr$. Zero time has been set at the moment when the
orbital parameters of the exact trajectory change their course abruptly,
attaining the values consistent with those reported for the S2 star. We
also show the graph of inclination vs.\ eccentricity (bottom right) with
an elliptical arc segment corresponding to Kozai's $c=0.275$. The
transition to a different regime of Kozai's oscillations occurs along
this segment and a detailed subplot of the indicated rectangular area
is enlarged in Fig.~\ref{fig2}.}
\label{fig1}
\end{figure*}

As mentioned above, the main difference between the direct numerical 
integration of stellar orbits and the analytical approach of the
averaging technique lies in the way how the rapidly changing (fast) 
variables are treated. The mean anomaly plays the role of a fast 
variable in our problem, therefore, the aim is to find a
canonical transformation under which the transformed Hamiltonian
becomes independent of this quantity. The transformation depends on the
form of the disturbing potential $V_\mathrm{d}(R,z)$, and so it cannot be
given explicitly for an arbitrary profile of density in the disc. One
can however introduce a formal development with
respect to the small parameter $\epsilon\ll1$. After transformation and
expansion to the first order we find
\begin{equation}
V_\mathrm{d}\rightarrow{\bar V}_\mathrm{d}={\bar V}_1+{\cal O}(\epsilon).
\label{vd}
\end{equation}
The first term of the series can be written explicitly,
\begin{equation}
\bar{V}_1=\frac{1}{2\pi\eta}\int_0^{2\pi}
\left(\frac{r}{a}\right)^2 V_{\rm{d}}({R},z)\;{\rm{d}}v,
\label{v1}
\end{equation}
where integration is performed over the true anomaly $v$ of the star for
one revolution around the black hole. At this level of approximation the
mean semimajor axis stays constant and differs from the osculating
semimajor axis by short-period terms only (we therefore do not
distinguish them in our notation). The problem is reduced to the
evolution of the mean eccentricity $e$ and pericentre argument $\omega$, 
which are further constrained by
$\bar{V}_1(e,\omega;a,c)=\mathrm{const}$.  When evaluating this
condition, $a$ and $c$ are parameters.

Levels of $\bar{V}_1=\mathrm{const}$ are shown in Figure~\ref{fig0}. We
use non-singular Poincar\'e coordinates $k(e,\omega)$ and $h(e,\omega)$,
defined by $h^2+k^2=2\left(1-\eta\right)$ and $h/k=\tan\omega$. In order
to draw these contours we construct a polar graph in $(\rho,\omega)$ 
with the azimuthal angle $0\leq\omega<2\pi$, polar radius 
$\rho\,\equiv\,\sqrt{h^2+k^2}$ and the origin $\rho=0$ coinciding with
the circular orbit (the mean eccentricity increases with polar radius,
$e\propto\rho$ near origin). The disc surface density was assumed 
constant.

Rigorous proof of convergence of the expansion (\ref{vd}) and an
estimation of accuracy of the first term $\bar{V}_1$ in the series is a 
difficult task which cannot be assessed for a general perturbation
potential. However, a partial check can be achieved by comparing the
analytical result with numerical integration of exemplary trajectories
(we will show such a comparison in the next section). Typically,
results based on a truncated part of the series are valid over a limited
time span and they fail to predict the system evolution for infinite
duration.

We conclude this section by summarizing the main points in our
perspective of how Kozai's phenomenon could help to solve the problem of
S-stars in Sgr~A$^\star$: 
\begin{description}
\item[(i)]~Axially symmetric (non-spherical) perturbation
of the central gravitational field is the essential element of the
proposed scheme. We attribute its origin to self-gravity of the gaseous
environment with the geometry of the disc.
\item[(ii)]~Stars of the nuclear cluster follow nearly Keplerian
orbits around the central black hole. Free motion in the central field
is disturbed by the disc gravity. 
\item[(iii)]~Star--disc collisions lead to dissipation and gradual
sinking toward the black hole, the decay rate being faster for
counter-rotating and highly eccentric trajectories. Eccentricity of
highly inclined orbits is pumped to large values via Kozai's mechanism.
Consequently, the pericentre distance shrinks.
\end{description}

\section{Results}
Examples of integration of the evolving trajectory are shown in 
Figure~\ref{fig1}, where the orbital oscillations and the progressive 
sinking of the star to the centre are evident. We introduce rescaled
time $\tau$ as a convenient measure, related to time $t$ in physical
units by 
\begin{equation}
t = \Sigma_3\,\left(\frac{M_\bullet}{3.5\times10^6M_{\sun}}\right)^2\tau.
\end{equation}
The characteristic time-scale $t_\mathrm{s}$ of the oscillations depends
on $M_\mathrm{d}/M_\bullet$ ratio and is generally much longer that the
dynamical time at the corresponding radius ($t_\mathrm{s}$ is of the
order of thousands of orbital revolutions in our case). On the other
hand, the rate of orbital decay $t_\mathrm{c}$ is proportional to 
$\Sigma_\ast$ as indicated by eq.~(\ref{tc}), and so our results can be
readily scaled to different stellar types.  Furthermore, scaling can be
performed to different values of $M_{\bullet}$ provided that the ratio
$M_\mathrm{d}/M_\bullet$ is retained, as well as the disc dimensions and
the orbital semimajor axis in units of $R_\mathrm{g}$. 

\begin{figure}
\includegraphics[width=\columnwidth]{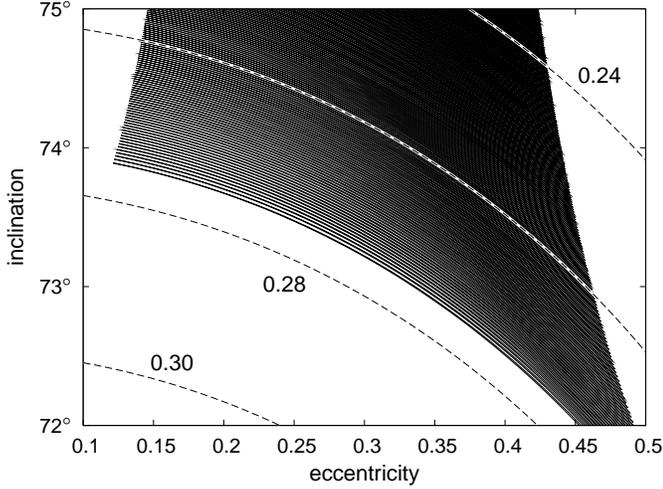}
\caption{Detail of the trajectory from Fig.~\ref{fig1}, showing that the 
orbit (solid line) oscillates almost precisely along the lines of
Kozai's integral (several contours are plotted with dashed lines and the
corresponding values of $c=\mathrm{const}$ are given). This indicates that the disc
mass is small but non-zero, and its gravity provides a perturbation of the
central field of the black hole. The orbit diffuses across the
$c$-curves due to orbital decay by dissipation.}
\label{fig2}
\end{figure}

\begin{figure}
\centering
\includegraphics[width=\columnwidth]{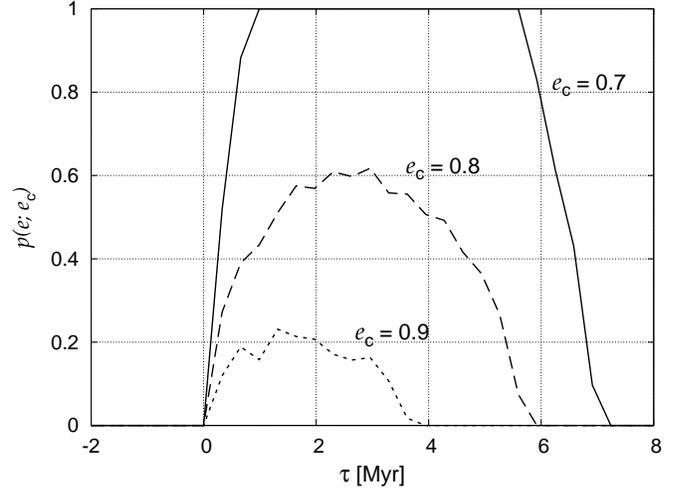}
\caption{
The probability $p(\tau;e_\mathrm{c})$ of attaining
eccentricity higher than $e_\mathrm{c}$ for the same example orbit as
in Figure~\ref{fig1}. Three curves are shown for different values of 
$e_\mathrm{c}$. The eccentricity reaches large values during
the period starting at $\tau=0$ and lasting about $6$~Myrs, and it
stays above $e_\mathrm{c}=0.7$ for a considerable fraction of time.}
\label{fig4}
\end{figure}

Now we set the inner edge of the disc at $R_\mathrm{in}=6R_\mathrm{g}$
and the outer edge $R_\mathrm{out}=5\times 10^4R_\mathrm{g}$. The disc
mass is $M_\mathrm{d}=10^{-3}M_\bullet$, which for $M_\bullet =
3.5\times10^6M_{\sun}$ gives
$\Sigma_\mathrm{d}=2.9\times10^{3}\mathrm{g\,cm}^{-2}$  for surface
density. In this case, Fig.~\ref{fig1} demonstrates how eccentricity is
pumped to large values and, consequently, the orbital decay is
accelerated. The semimajor axis decreases by a factor of 5 in several
million years, even for a relatively moderate choice of the disc mass
and density. The effect is thus potentially important, as it is able to
bring stars near the black hole and put them on eccentric orbits. In
these computations we employed a fourth-order Runge-Kutta integrator
with adaptive stepsize. In order to reduce the duration of runs we
pre-generated the field components in a Cartesian grid covering a
rectangular region of 1~pc\,$\times$\,1~pc. In some regions, close to
the disc, the field lines are highly distorted. Therefore, we used
additional sub-grids that provide fine resolution where needed. Field
values in between mesh points were obtained by bilinear interpolation.
As detailed below, we can be fairly confident of sufficient accuracy of
the numerical computation because the resulting trajectories agree with
expectations based on analytical arguments of the perturbation method.

The orbital oscillations,
which we observe in graphs, call for some kind of probabilistic 
interpretation of the orbital parameters. This can be inferred from
non-uniform distribution of dots which represent fluctuations of $e(\tau)$
and $i(\tau)$ in Fig.~\ref{fig1}. Each dot represents the position in the
phase space at a corresponding moment of time, and the density of dots
determines the probability that the star occurs with given values of
parameters.\footnote{To avoid 
fake structures in the graph, dots are plotted with random time steps
that are distributed uniformly over $\langle0,t_\mathrm{max}\rangle$,
where $t_\mathrm{max}$ is an interval longer than the orbital time and
shorter than characteristic time $t_\mathrm{s}$ of orbital
oscillations.} From Figure~\ref{fig1} and especially from the detail in
Figure~\ref{fig2} we see that the oscillations of eccentricity and
inclination are bound together by Kozai's quasi-integral
$c\,\equiv\,\sqrt{1-e^2}\,\cos i=\mathrm{const}$.

\begin{figure*}
\includegraphics[width=\textwidth]{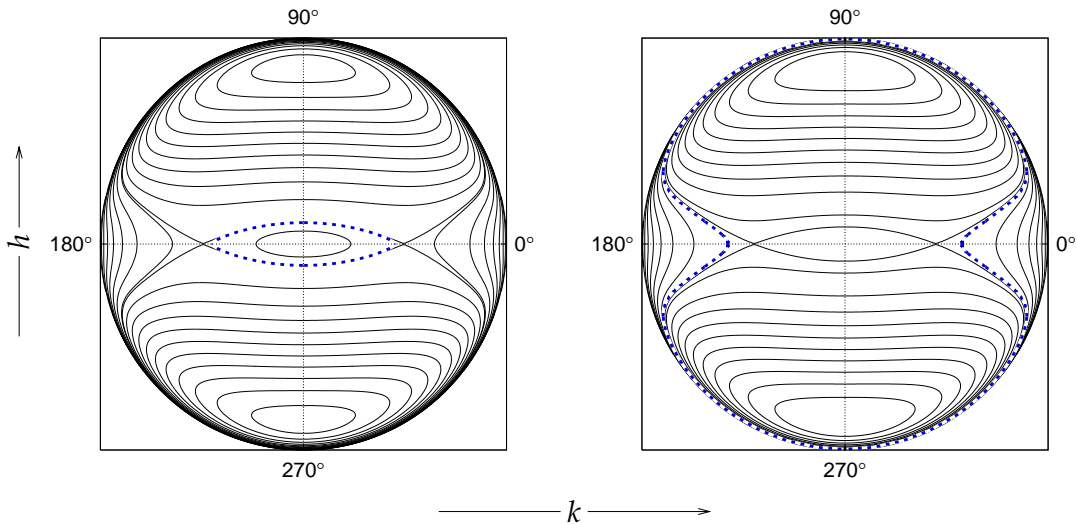}
\caption{The contour plot of disturbing potential
$\bar{V}_1(\rho,\omega)=\mathrm{const}$ shown at two moments of time. 
The exact (numerically integrated) trajectory is also plotted (by
points). The left panel corresponds to the moment just before the
numerical trajectory crosses the separatrix: $\tau=-0.15\mathrm{Myr}$,
$a=2.50\times10^4R_\mathrm{g}$. The corresponding value of Kozai's
$c=0.2733$ and the maximum eccentricity $e_\mathrm{max}=0.962$. On the
other hand, the right panel captures the orbit after crossing the
separatrix: $\tau=0.15\mathrm{Myr}$, $a=2.46\times10^4R_\mathrm{g}$,
$c=0.2795$, $e_\mathrm{max}=0.960$.}
\label{fig3}
\end{figure*}
Another representation of the orbital oscillations is given in 
Figure~\ref{fig4} where we plot the time evolution of probability
$p(\tau;e_\mathrm{c})$ that eccentricity has exceeded a given threshold
$e_\mathrm{c}$. This graph demonstrates how a sinking star can, indeed,
be set on a highly elliptic trajectory at certain stages, while
gradually approaching the black hole. Fig.~\ref{fig4} helps to answer
the question of what fraction of time the orbit actually spends at the
stage of high eccentricity. For example, by integrating
${\int}p(\tau;e_\mathrm{c}=0.8)\,\mathrm{d}\tau$ over the first
$6$~Myrs we find the star spends approximately half its time with an
eccentricity larger than $0.8$. Then the orbit becomes progressively
circularized and the star descends below $10^4R_\mathrm{g}$.

Contour levels of the disturbing potential are shown in
Figure~\ref{fig3}. We construct this graph using the same procedure as
described for Fig.~\ref{fig0}. Two panels
correspond to slightly different moments; notice that there is a small
change of the contour line structure because of continuous dissipation.
In these plots we also include the exact (numerically integrated)
trajectory. Clearly the adopted approximation is sufficiently accurate:
the numerical trajectory follows its underlying contour at each
corresponding panel, as it should. An adiabatic diffusion across
$\bar{V}_1=\mathrm{const}$ lines is slow and it cannot be resolved in this
figure, where $\omega$ circulates over the whole interval
$\langle0,2\pi\rangle$ on the time-scale of the order of several hundred
orbital periods. However, there is an evident difference between the two
cases: in the left panel the trajectory oscillates near the origin keeping
small eccentricity, while on the right it has crossed the separatrix. As
a result of this change, the corresponding orbital energy is also
slightly different, whereas the mean eccentricity starts to vary over
substantially larger intervals. This also explains how the averaged
model can be used to estimate the maximum eccentricity attained by the
star.

In the above example the Kozai oscillations were only used to set the S2
orbit into a highly eccentric state. Indeed, it is evident from the
orbit evolution at $\tau>0$ that its decay is considerably accelerated
by this process. This is due to the fact that a star on a highly
eccentric orbit passes through the disc with large relative velocity.
The same argument suggests that counter-rotating stars suffer more from
the orbital dissipation than co-rotating ones. Moreover, the efficiency of
dissipation due to large inclination is strengthened because the
separatrix prevents the overturning of counter-rotating orbits to
co-rotation.

\begin{figure*}
\includegraphics[width=\textwidth]{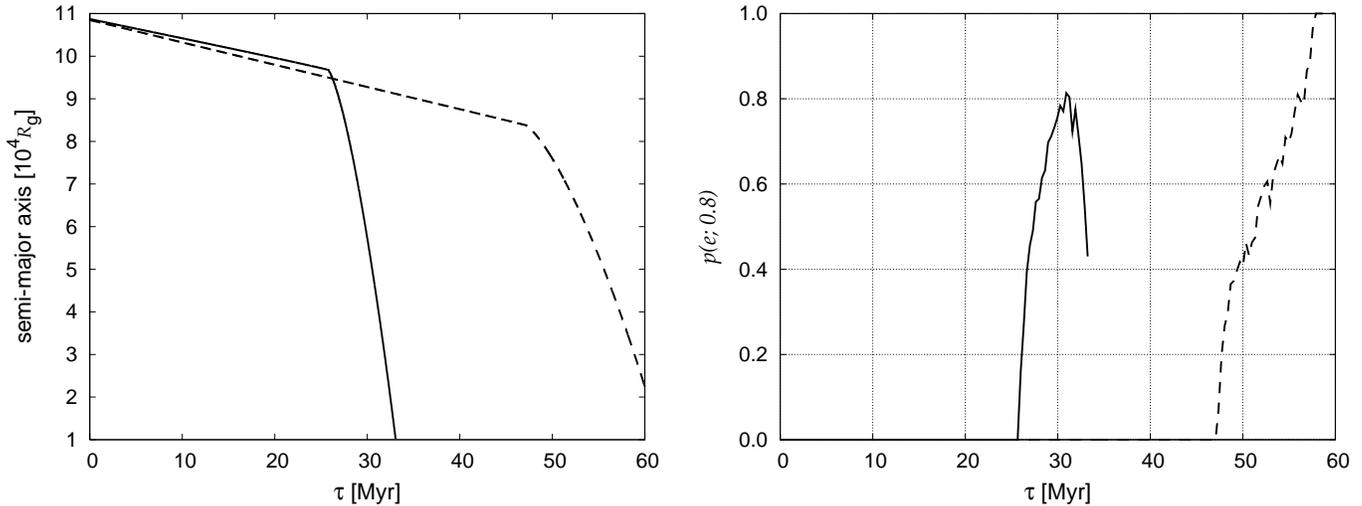}
\caption{Left: The orbital decay due to interaction with the disc.
Two cases are shown corresponding to a homogeneous disc 
$\Sigma_\mathrm{d} \propto \mathrm{const}$ (solid line), and
$\Sigma_\mathrm{d} \propto \sqrt{R(R_\mathrm{out}-R)}$ (dashed line) 
representing an annulus. The rate of gas-assisted orbital  decay speeds
up at the moment of crossing the separatrix (time $\tau=25$~Myrs and
$\tau=47$~Myrs, respectively). Right: the probability for these orbits to
have eccentricity larger than $0.8$.  Both trajectories start with
identical initial values, $a_\mathrm{i} = 1.1\times
10^5R_\mathrm{g},\,\,e_\mathrm{i}=0.15$ and $i_\mathrm{i} = 105\degr$,
and at some stage they attain the values compatible with the S2-star
parameters, i.e.\ $a\approx2.5\times10^4R_\mathrm{g}$ and
$e\approx0.87$.}
\label{fig_fast_decay}
\end{figure*}

Hence, we found it convenient to explore the possibility that the
S-stars have been formed on nearly circular orbits farther form the
centre. Then, due to the simultaneous energy dissipation and the
influence of the perturbing gravitational field the stars undergo orbital
decay. Sample trajectories in Figure~\ref{fig_fast_decay} were
integrated assuming $R_\mathrm{out}=2\times 10^5R_\mathrm{g}$ and
$M_\mathrm{d}/M_\bullet=0.01$. Two different lines represent the orbital
decay of a star interacting with different $\Sigma_\mathrm{d}(R)$
profiles starting with identical initial values of orbital elements.

The region of efficient star formation is usually placed at
$r\approx10^6R_\mathrm{g}$, however, the minimum radius is not known
very precisely. Collin \& Zahn (\cite{collin99}) argue, using the model
of a marginally unstable self-gravitating accretion disc, that the star
forming process can take place at distances $\gtrsim10^5R_\mathrm{g}$.
Although their work was intended for quasars, it indicates that the
mechanism discussed here could be relevant also for Sgr~A$^\star$ if
there is enough material dispersed around the black hole, or if it was
there in the past. Star forming process occurring at  roughly
$10^5R_\mathrm{g}$ was also considered by Nayakshin \& Cuadra
(\cite{nay04a}). Furthermore, it was suggested (Genzel et al.\
\cite{genzel03}) that young He~{\sc{}i} stars rotating in two coherent
rings of radii $\gtrsim 10^5R_\mathrm{g}$ could have formed during a recent star
forming episode triggered by a collision of different gaseous streams.
Our two examples in Figure~\ref{fig_fast_decay} indicate that the exact
form of the gas distribution play an essential role, both for the orbital
decay and for oscillations. Fine-tuning of the model parameters changes
the quantitative results and allows us to decrease $M_\mathrm{d}$ but
these details cannot be compared with present-day observations. 

As explained above, jumps of mean eccentricity are caused by the Kozai
process in an non-spherical (axially symmetric) perturbation of the
central gravitational field. This we interpret as the influence of the
disc, which simultaneously provides the dissipative environment in a
self-consistent manner. The process of Kozai-driven resonance prefers
rather special orientation of initial trajectories, namely, almost
perpendicular inclination. We can therefore speculate that the current
family of S-stars was born jointly and these were originally moving in a
similar direction. They experienced the Kozai resonance and underwent
the orbital decay ending in the present state of highly elliptic
trajectories very near the black hole. Some of the nuclear stars in
Sgr~A$^{\ast}$ (Genzel et al.\ \cite{genzel03}; Levin \& Beloborodov
\cite{levin03}) form a disc-like structure rather than a spherical
distribution (nuclear stars are actually located in two rings, almost
perpendicular to each other; Genzel et al.\ \cite{genzel03}). We can
thus consider such rings as current prototypes of a formerly existing
ring in which S-stars were formed.

\section{Discussion}
\label{sect:discussion}
In this paper we considered two complementary effects which both act as
perturbations on otherwise free Keplerian motion, namely, gravity of the
disc and the orbital decay due to crossing the disc slab. We computed a
set of model trajectories of stars sinking towards a supermassive black
hole in Sgr~A$^{\ast}$. We considered the region of gravitational
dominance of the central black hole, however, we also took the gravity
of the accretion disc and the corresponding dissipation into account.
Stars do not strictly follow test-body (geodesic) motion in the central
field, but instead they gradually sink to the centre and they experience
occasional jumps of mean orbital elements. These second-order effects do
not influence the motion on short time-scales of the order of one or
several revolutions, nevertheless they are essential for long
time-scales. Although the existence of a fossil disc in the Galaxy
centre is a matter of debate, the idea offers a natural explanation for
various pieces of observational information, such as the occurrence of
X-ray flares or the origin of young OB-type stars on highly elongated
orbits. We therefore adopted the S2 orbital parameters in our
simulations.

A more complete picture will require not only a detailed knowledge of
the disc structure but also coupling with other effects, such as
dynamical friction, gravitational radiation losses and the internal
dissipation by tidal forces, which also contribute to the orbital decay,
as well as the Lense-Thirring precession caused by rotation of the black
hole. These and other mechanisms are beyond the scope of the present
paper. Likewise, we neglected the effect of grazing encounters with
other stars in the nuclear cluster. All these processes will have to be
taken into account in order to disentangle their impact on the evolving
orbit and to measure black hole parameters in the future. Our
model predicts that stars spend a considerable fraction of total time 
at the stage of large eccentricity. This can be reconciled with the current 
evidence which strongly prefers high eccentricities in the central arcsecond
region of Sgr~A$^\star$, because highly elliptical orbits are
preferentially brought to small radii. 

The system of stars on energetically bound trajectories encircling a
supermassive black hole is a unique configuration allowing one to probe
the gravitational field of the black hole by tracking the stellar proper
motion. The available accuracy has justified the approximation of free
motion, assuming that the stars behave as test particles that are
insensitive to non-gravitational forces. This is fully satisfactory on
the time-scale of several revolutions, i.e., of the order of tens to
hundreds of years. However, we showed that the effect of gaseous and/or
dusty environment should not be neglected in the discussion of the
long-term (Myr) evolution and that it can in fact help to reconcile the
problem of the origin of the stellar population in the Galaxy centre.
Although our approach involves various simplifications, the main message
here is that one should take both effects into account simultaneously,
i.e.\ the orbital dissipation in the gaseous environment of the disc and
its gravitational field. In fact, the idea of combining gravitational
(conservative) and hydrodynamical (dissipative) star--disc interactions
can be considered as a model case for a more general kind of process
introducing a non-spherical perturbation to the central gravitational
field and, simultaneously, a dissipative mechanism for the orbital
decay.

\begin{acknowledgements}
We thank S.~Collin for valuable discussions about star formation in
black-hole accretion discs with self-gravity, and to K.~Beckwith,
C.~Hopman, N.~Mouawad and D.~Vokrouhlick\'y for comments. We also thank
our referees, S.~Nayakshin and J.~Cuadra, for helpful criticism
improving the paper. We gratefully acknowledge the financial support
from the Czech Science Foundation under grants 205/02/P089 (L\v{S}) and
205/03/0902 (VK), as well as Charles University in Prague (299/2004).
\end{acknowledgements}


\begin{thebibliography}{}
\bibitem[2003]{alex03}Alexander~T., Hopman~C., 2003, 560, L143
\bibitem[2001]{alex01}Alexander~T., Livio~M., 2001, 590, L29
\bibitem[2004]{alex04}Alexander~T., Livio~M., 2004, 606, L21
\bibitem[1996]{arm96}Armitage P.~J., Zurek W.~H., Davies M.~B., 1996, ApJ, 470, 237
\bibitem[1989]{arnold89}Arnold V.~I., 1989, Mathematical Methods of Classical Mechanics (Springer-Verlag: Berlin)
\bibitem[1993]{art93}Artymowicz~P., 1993, ApJ, 419, 166
\bibitem[1993]{art93b}Artymowicz~P., Lin D.~N.~C., Wampler E.~J., 1993, ApJ, 409, 592
\bibitem[2004]{asch04}Aschenbach~B., Grosso~N., Porquet~D., Predehl~P., 2004, A\&A, 417, 71
\bibitem[2001]{bag01}Baganoff~K., Bautz~W., Brandt~N.; Chartas~G.\ et al., 2001, Nature, 413, 45
\bibitem[2002]{blaes02}Blaes~O., Lee M.~H., Socrates~A., 2002, ApJ, 578, 775
\bibitem[1961]{brower61}Brower~D., Clemence~G., 1961, Methods of Celestial Mechanics (Academic Press: New York)
\bibitem[1999]{collin99}Collin~S., Zahn J.-P., 1999, A\&A, 344, 433
\bibitem[2003]{cuadra03}Cuadra~J., Nayakshin~S., Sunyaev~R., 2003, A\&A, 411, 405
\bibitem[2004]{eckart04}Eckart~A., Moultaka~J., Viehmann~T., Straubmeier~C., Mouawad~N., 2004, ApJ, 602, 760
\bibitem[2003]{genzel03}Genzel~R., Sch\"odel~R., Ott~T., Eisenhauer~F.\ et al., 2003, ApJ, 594, 812
\bibitem[2003a]{ghez03a}Ghez A.~M., Becklin E.~E., Duch\^ene~G., Hornstein~S., Morris~M., Salim~S., Tanner~A., 2003a, Astron. Nachr., 324, 527
\bibitem[2003b]{ghez03b}Ghez A.~M., Duch\^ene~G., Matthews K., Hornstein S.~D.\ et al., 2003b, ApJ, 586, L127
\bibitem[2004]{goodman04}Goodman~J., Tan J.~C., 2004, ApJ, 608, 108
\bibitem[1994]{ham94}Hameury J.-M., King A.~R., Lasota J.-P., Auvergne~M., 1994, A\&A, 292, 404
\bibitem[2003]{han03}Hansen B.~M.~S., Milosavljevi\'c~M., 2003, ApJ, 593, L80
\bibitem[1997]{holman97}Holman~M., Touma~J., Scott~T., 1997, Nature, 386, 254
\bibitem[2001]{karas01}Karas~V., \v{S}ubr~L., 2001, A\&A, 376, 686
\bibitem[1994]{karas94}Karas~V., Vokrouhlick\'y, 1994, ApJ, 422, 208
\bibitem[1999]{kino99}Kinoshita~H., Nakai~H., 1999, Celest. Mech., 75, 125
\bibitem[1998]{kis98}Kiseleva L.~G., Eggleton P.~P., Mikkola~S., 1998, MNRAS, 300, 292
\bibitem[1962]{koi62}Kozai~Y., 1962, AJ, 67, 591
\bibitem[1983]{lb83}Lass H., Blitzer L., 1983, Celest. Mech. Dyn. Astron., 30,~225
\bibitem[2003]{levin03}Levin~Y., Beloborodov A.~M., 2003, ApJ, 590, L33
\bibitem[1975]{lidov76}Lidov M.~L., Ziglin S.~L., 1976, Celest. Mech., 13, 471
\bibitem[2003]{mcm03}McMillan S.~L.~W., Portegies Zwart S.~F., 2003, ApJ, 596, 314
\bibitem[2004]{mil04}Milosavljevi\'c~M., Loeb~A., 2004, ApJ, 604, L45
\bibitem[2002]{morbi02}Morbidelli~A., 2002, Modern Celestial Mechanics (Taylor \& Francis: London)
\bibitem[1993]{morris93}Morris~M., 1993, ApJ, 408, 496
\bibitem[2004]{mou04}Mouawad~N., Eckart~A., Pfalzner~S., Sch\"odel~R., Moultaka~J., Spurzem~R., 2004, A\&A, submitted (astro-ph/0402338)
\bibitem[2004a]{nay04a}Nayakshin~S., Cuadra~J., 2004a, A\&A, submitted (astro-ph/0409541)
\bibitem[2004b]{nay04b}Nayakshin~S., Cuadra~J., Sunyaev~R., 2004b, A\&A, 413, 173
\bibitem[2003]{nay03}Nayakshin~S., Sunyaev~R., 2003, MNRAS, 343, L15
\bibitem[2004]{pierens04}Pierens~A., Hur\'e J.-M., 2004, ApJ, 605, 179
\bibitem[1995]{rauch95}Rauch~K., 1995, MNRAS, 275, 628
\bibitem[2003]{schodel03}Sch\"odel~R., Ott~T., Genzel~R., Eckart~A., Mouawad~N., Alexander~T., 2003, ApJ, 596, 1015
\bibitem[2002]{schodel02}Sch\"odel~R., Ott~T., Genzel~R., Hofmann~R.\ et al., 2002, Nature, 419, 694
\bibitem[1999]{sem99}Semer\'ak~O., Karas V., de Felice F., 1999, PASJ, 51, 571
\bibitem[1999]{sridhar99}Sridhar~S., Touma~J., 1999, MNRAS, 303, 483
\bibitem[1999]{subr99}\v{S}ubr~L., Karas~V., 1999, A\&A, 352, 452
\bibitem[2004]{subr04}\v{S}ubr~L., Karas~V., Hur\'e J.-M., 2004, MNRAS, 354, 1177
\bibitem[1991]{syer91}Syer~D., Clarke C.~J., Rees M.~J., 1991, MNRAS, 250, 505
\bibitem[1996]{thomas96}Thomas~F., Morbidelli~A., 1996, Celest. Mech. Dyn. Astron., 64, 209
\bibitem[2002]{vilk02}Vilkoviskij E.~Y., Czerny~B., 2002, A\&A, 387, 804
\bibitem[1993]{vok93}Vokrouhlick\'y~D., Karas~V., 1993, MNRAS, 265, 365
\bibitem[1998a]{vok98}Vokrouhlick\'y~D., Karas~V., 1998a, MNRAS, 293, L1
\bibitem[1998b]{vok98b}Vokrouhlick\'y~D., Karas~V., 1998b, MNRAS, 298, 53
\bibitem[2003]{wen03}Wen~L., 2003, ApJ, 598, 419
\bibitem[1983]{zen83}Zentsova A.~S., 1983, Ap\&SS, 95, 11
\bibitem[1994]{zurek94}Zurek W.~H., Siemiginowska~A., Colgate S.~A., 1994, ApJ, 434, 46
\end{thebibliography}
\end{document}